\documentclass[reprint,amsmath,amssymb,aps]{revtex4-1}
\usepackage{lipsum}
\usepackage{xcolor}
\usepackage{graphicx}
\usepackage{dcolumn}
\usepackage{bm}
\usepackage{natbib}
\usepackage{enumitem}  
\usepackage{float}
\newcommand{\sgn}{\rm sign}
\begin{document}
\preprint{APS/123-QED}
\title{Functional renormalization group study of parallel double quantum dots:\\Effects of asymmetric dot-lead couplings}
\author{V. S. Protsenko}
\author{A. A. Katanin}%
\affiliation{%
M. N. Mikheev Institute of Metal Physics, 620990 Ekaterinburg Russia\\
Ural Federal University, 620002 Ekaterinburg, Russia
}
\date{\today}
\begin{abstract}
We explore the effects of asymmetry of hopping parameters between double parallel quantum dots and the leads on the conductance and a possibility of local magnetic moment formation in this system using functional renormalization group approach with the counterterm. 
We demonstrate a possibility of a quantum phase transition to a local moment regime (so called singular Fermi liquid (SFL) state) for various types of hopping asymmetries and discuss respective  gate voltage dependences of the conductance. It is shown, that depending on the type of the asymmetry, the system can demonstrate either a first order quantum phase transition to SFL state, accompanied by a  discontinuous change of the conductance, similarly to the symmetric case, or the second order quantum phase transition, in which the conductance is continuous and exhibits Fano-type asymmetric resonance near the transition point. A semi-analytical explanation of these different types of conductance behavior is presented.
\end{abstract}
\maketitle
\section{Introduction}\par
Nanostructures based on quantum dots attract growing interest due to an opportunity of 
tuning of 
their transport and magnetic properties \cite{Busl_2010,Zitko_2007(2),Tooski_2014,Mravlje_2006,Mitchell_2010,Wright_2011,Hsieh_2012}, which makes them 
promising candidates for quantum spintronics and quantum information processing applications \cite{Loss_1998,DiVincenzo_2005,Bennett_2000,Engel_2004,Awschalom_2002,Nielsen_2001,DiVincenzo_2000,Petta_2005}. At appropriate conditions, these systems 
consist of discrete energy levels of quantum dots, which are hybridized with the leads, having continuous energy bands.  It is well known that physical properties of such zero-dimensional structures are strongly influenced by Coulomb electron
interaction effects, leading to non-trivial interaction-induced effects \cite{Kouwenhoven_1997,Andergassen_2010,Reimann_2002} (e.g., the Kondo effect \cite{Hewson_1997}). 

On the other hand, a specific quantum dots arrangements, in which multipath propagation through system are possible, can give rise
to quantum interference effects \cite{Hackenbroich_2001,Guevara_2003,Guevara_2006,Gong}. The essential feature of these effects is appearance of resonance peak structures in the conductance, making electronic transport properties very sensitive to small changes of parameters, which may be important for
practical applications.
In this context, the interplay and cooperation between the correlation effects and quantum interference, associated with a system geometry, can be significantly important and provide unexpected 
electron transport features \cite{Meden_2006,Karrasch_2006,Trocha_2007,Trocha_2012,Oguri_2011}. 

Recently, it was found that quantum dot systems with ring geometries, realizing quantum interference effects in presence of interaction, may demonstrate the interaction-induced quantum phase transition to the so-called singular Fermi liquid (SFL) state, characterized by local magnetic moment in one of the effective (``odd") states \cite{Zitko_2006,Zitko_2007(2),Zitko_2007(1),Zitko_2012}. In particular, in the simplest ring geometry of the system, consisting of two quantum dots coupled in parallel to two common leads, the appearance of the phase transition to the SFL state is related to the specific electron redistribution between the even and odd states. 
For the parallel quantum dot system with all hopping parameters between dots and leads equal, the SFL state has been studied by various methods \cite{Zitko_2012,IILM} and was shown to appear due to the full decoupling of the odd state from the leads, which yields formation of the local magnetic moment in the system. 

Although the 
electronic transport 
in various models of asymmetric parallel double quantum dot systems (e.g., non-interacting~\cite{Guevara_2003,Gong} and with Coulomb interactions~\cite{
Meden_2006,Tanaka_2005,Sztenkiel_2007, 
Wong_2012,Dias_da_Silva_2006,Trocha,Trocha_2007,
Karrasch_2006,Zitko_2006}) was studied earlier, the effect of asymmetry on SFL state remains not fully investigated. For strong Coulomb interaction it was suggested \cite{Wong_2012} that in the presence of weak asymmetry of interactions on quantum dots the formation of the spin-half SFL state is realized with decreasing temperature via the underscreened Kondo effect \cite{Posazhennikova}. At the same time, 
the effect of the asymmetry of dot-leads hopping parameters (which is unavoidably present in the experimental setups) on the presence of the local moments and the possibility of realization of SFL state, especially for 
not too strong Coulomb interaction,
was not investigated in detail. 
\par
Numerical efforts, which are necessary for the 
existing numerical methods (e.g., numerical renormalization group (NRG) \cite{Bulla_2008}, quantum Monte Carlo 
\cite{Hirsch_1986,Fye_1988}, continuous-time quantum Monte Carlo 
\cite{Gull_2011}, exact diagonalization 
\cite{Dagotto,Torio_2002,Busser_2005}, nano-DMFT \cite{nanoDMFT1,nanoDMFT2,nanoDMFT4_1,nanoDMFT3,nanoDMFT4,nanoDMFT5}, and nano-D$\Gamma$A \cite{nanoDMFT4_1,nanoDGA}) grow fast with increasing system size or asymmetry, 
such that the comprehensive analysis of 
complex quantum dot systems (especially the conductance) is rather difficult for purely numerical methods. Therefore, developing and using semi-analytical techniques is important for description of such systems. 

One of promising methods, which mostly overcomes the above discussed numerical difficulties and has been successfully applied for investigating the effects of electron interaction in different nanoscopic systems is the functional renormalization group (fRG) method \cite{Metzner_2012,Salmhofer_2001}. This method results in an exact hierarchy of differential flow
equations for the irreducible vertex functions (self-energy, effective two-particle interaction and higher-order irreducible vertices). With a suitable truncation fRG equations can be reduced to a closed set and then can be easily integrated numerically. 
The implementation of the fRG approach is rather flexible comparing to the existing numerical methods; this method recently has been formulated in the Keldysh formalism (see, e.g., Refs.~\cite{Gezzi_2007,Jakobs_2007}) and on the real time axis \cite{Meden_2012}, which makes it applicable to different non-equilibrium problems (e.g., considering a finite bias voltage or a real time evolution).\par
Although this method was adapted to study quantum dot systems, including fairly complex geometries, long time ago \cite{Andergassen_2010,Meden_2006,Karrasch_2006,Meden_2008}, 
only recently its modification, allowing to describe SFL state and  providing a good agreement with the numerical renormalization group data for parallel quantum dot system up to the intermediate value of the interaction, was proposed \cite{IILM}.

In the present paper, using this latter approach, we present a systematic study of the effects of an  asymmetric coupling of parallel quantum dots to the leads on electron transport and local magnetic moment formation, yielding a possibility of realizing the SFL state. 
We find that formation of the local magnetic moment in parallel double quantum dot system takes place for a wide range of asymmetries. We also clarify what features can be observed in the gate voltage dependence of the zero temperature linear conductance in the limit of zero magnetic field for each class of asymmetry of hybridizations and their effect on quantum phase transition. We show, that for some types of asymmetry the asymmetric Fano-like resonance is formed in the linear conductance. Finally, a semi-analytical explanation of the observed features for arbitrary asymmetry is provided.\par
This paper is organized as follows. In Sect. II we introduce the model and briefly formulate the counterterm extension of fRG method.  In Sect. III we 
present the fRG results for the conductance and analyze a possibility of the local moment formation.
Finally, in Sect. IV we present conclusions. 
\section{Model and method}\par
\begin{figure}[h]
\center{\includegraphics[width=0.55\linewidth]{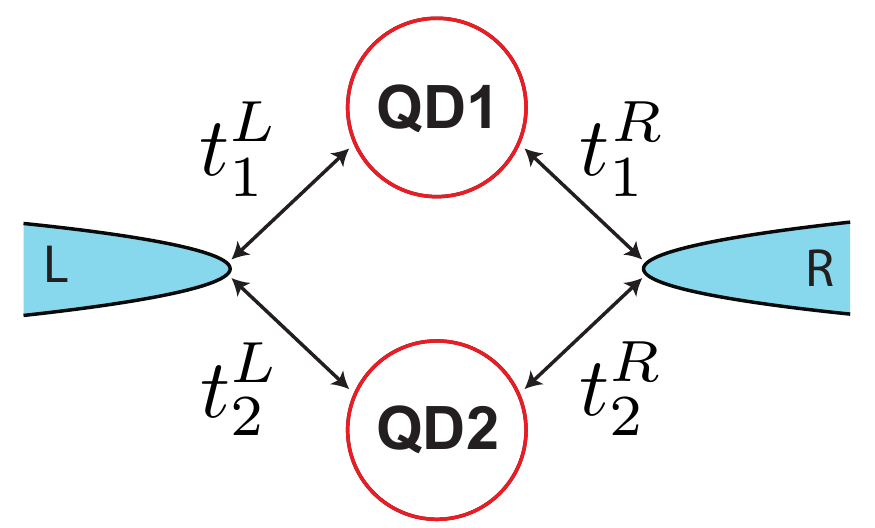}}
\caption{(Color online). The sketch of the quantum dot system.}
\label{sketch}
\end{figure}

We consider a system (see Fig.~\ref{sketch}), consisting of two single-level quantum dots, QD1 and QD2, coupled in parallel to two, left (L) and right (R), non-interacting leads.\par
The Hamiltonian of the system is given by
\begin{equation}
 \mathcal{H}= \mathcal{H}_{\rm dot}+\mathcal{H}_{\rm lead}+\mathcal{H}_{\rm coupl}.
 \label{Hamiltonian}
\end{equation}
The first term represents the Hamiltonian of isolated quantum dots
\begin{equation}
 \mathcal{H}_{\rm dot}=\sum_{\sigma}\sum_{j=1}^{2}\left[\left(\epsilon_{\sigma}-\dfrac{U} {2}\right)n_{j,\sigma}+\dfrac{U}{2}n_{j,\sigma}n_{j,\bar{\sigma}}\right],
 \label{H_dot}
\end{equation}
where $n_{j,\sigma}=d^{\dagger}_{j,\sigma}d_{j,\sigma}$ denotes the electron number operator, with creation (annihilation) operators $d^{\dagger}_{j,\sigma}(d_{j,\sigma})$ for an electron with spin projection $\sigma=\pm {1}/{2}$
and $\bar{\sigma}=-\sigma$ on quantum dot $j=\{1,2\}$, $\epsilon_{\sigma}$ and $U$ denote the level position and the on-site Coulomb interaction,  respectively. The level position $\epsilon_{\sigma}$ can be changed by applying of the gate voltage $V_{g}$ and magnetic field $H$, thus $\epsilon_{\sigma}=V_g-\sigma H$.  The leads are modeled by
\begin{equation}
\mathcal{H}_{\rm lead}=-\tau\sum_{\alpha=L,R}\sum_{j=0}^{\infty}\sum_{\sigma}(c^{\dagger}_{\alpha,j+1,\sigma}c_{\alpha,j,\sigma}+\text{H.c.}),
\label{H_lead}
\end{equation}
where $c^{\dagger}_{\alpha,j,\sigma}(c_{\alpha,j,\sigma})$ is the corresponding creation (annihilation) operator and $\tau$ denotes  nearest-neighbor hopping between the sites of the leads.
Finally, the coupling between quantum dots and the leads is given by 
\begin{equation}
\mathcal{H}_{\rm coupl}=-\sum_{\alpha=L,R}\sum_{j}\sum_{\sigma}(t^{\alpha}_{j}c^{\dagger}_{\alpha,0,\sigma}d_{j,\sigma}+\text{H.c.}),
\label{H_coupl}
\end{equation}
where $t^{\alpha}_{j}$ is the hopping matrix element between lead
$\alpha$ and $j$--th quantum dot.\par

{\it Method.}
To treat the effects of two particle interaction $U$, we use the one-particle irreducible (1PI) version of the fRG method \cite{Metzner_2012}, supplemented by the counterterm, recently introduced in Ref. \cite{IILM}, which allows for treatment of local moments. This method starts with considering the non-interacting propagator
of quantum dot system, obtained by projection of the leads and taking the wide-–  band limit \cite{Karrasch_2006,Karrasch_thesis,Enss_thesis}, as a matrix in the quantum dot space \cite{Karrasch_2006},
\begin{eqnarray} 
\mathcal{G}_{0,\sigma}^{-1}(i\omega)&=&\left(i\omega-\epsilon_{\sigma}+U/2\right)\mathbf{I}\notag\\
&+&i\begin{pmatrix}
\Gamma^{L}_{1}+\Gamma^{R}_{1} & \Gamma_{12}\\
\Gamma_{12}&\Gamma^{L}_{2}+\Gamma^{R}_{2}
\end{pmatrix}\sgn(\omega),
\label{Green_function}
\end{eqnarray}
where $\Gamma^{\alpha}_{j}=\pi|t^{\alpha}_{j}|^{2}\rho_{\text{lead}}(0)$ denotes the energy independent hybridization strength in the wide-band limit of the leads, $\rho_{\text{lead}}$ is the local density of the states of the leads, $\Gamma_{12}=\sum_{\alpha}{\left(\Gamma^{\alpha}_{1}\Gamma^{\alpha}_{2}\right)^{1/2}}$, and ${\bf I}$ denotes identity matrix in the quantum dot space. 

To construct the fRG flow this non-interacting propagator is replaced by a flow parameter $\Lambda$ dependent one, such that $\mathcal{G}^{\Lambda=\Lambda_0}_{0,\sigma}(i\omega)=0$ corresponds to the non-interacting problem, while $\mathcal{G}^{\Lambda=0}_{0,\sigma}(i\omega)=\mathcal{G}_{0,\sigma}(i\omega)$ corresponds to the problem studied, $\Lambda_0$ is the initial value of the parameter $\Lambda$. Specifically, we choose 
\cite{IILM}
\begin{equation}
\mathcal{G}^{\Lambda}_{0,\sigma}(i\omega)=\left[\mathcal{G}^{-1}_{0,\sigma}(i\omega)+f^{\Lambda}(\omega)+\chi_{\sigma}^{\Lambda} \right]^{-1},
\label{cutoff}
\end{equation}
where second term $f^{\Lambda}$ in the square brackets of Eq.~(\ref{cutoff}) regulates (fermionic) infrared modes of the bare propagator. We use the Litim-type regulator \cite{Litim_2001} of the form \cite{IILM} 
$$
f^{\Lambda}(\omega)=i{\bf I}\left(\Lambda-|\omega|\right)\Theta\left(\Lambda-|\omega|\right)\sgn(\omega),
$$
which as shown in Ref.~\cite{IILM} somewhat improves the results of the standard fRG scheme with the sharp cutoff.\par 
The last term $\chi_{\sigma}^{\Lambda}$ in Eq. (6) is a counterterm, which serves as an infrared regulator in the two-particle sector, and, as shown in the previous paper \cite{IILM},  eliminates the 
problem of the divergences of the vertices in the fRG flow, allowing to describe the singular Fermi liquid (SFL) phase of the system. The counterterm provides switching on/off of the additional magnetic field $\tilde{H}$ at the beginning ($\Lambda=\Lambda_{0}$)/end ($\Lambda\rightarrow 0$) of the fRG flow and 
is chosen to have linear dependence on the cutoff parameter of the form
\begin{equation}
\chi_{\sigma}^{\Lambda}=\sigma\tilde{H} \min (1,\Lambda/\Lambda_{c}){\bf I}.
\end{equation}
The parameter $\Lambda_{c}$  in the above equation determines sharpness of switching off the additional field $H_{c}$ and chosen according to the value of this field \cite{IILM}.

After differentiation generating
functional of the irreducible vertex functions with respect to $\Lambda$, one obtains an infinite hierarchy of differential equations for the $n$--particle vertex functions $\Gamma^{n}$. In the present study we, following Ref.~\cite{Karrasch_2006}, truncate the fRG equations by discarding the contribution of the vertices with $n>2$ and neglect frequency dependence of the one--particle (self-energy $\Sigma^{\Lambda}$) and two--particle (effective two particle interaction $\Gamma^{\Lambda}$) vertices. This approximation was shown to describe well the electronic and transport properties of single- and multiple quantum dot systems \cite{Karrasch_2006,IILM}. In particular, it allows describing the singular Fermi-liquid state, which appears because of the disconnection of the odd level from the leads \cite{IILM}. In this way we obtain a closed set of standard fRG equations of the form \cite{Karrasch_2006}
\begin{eqnarray}
\partial_{\Lambda}\Sigma^{\Lambda}_{m^{'}m}&=&-\int\dfrac{d\omega}{2\pi}\mathcal{S}^{\Lambda}_{n n^{'}}\Gamma^{\Lambda}_{m^{'}n^{'}m n}, \label{Gamma} \\
\partial_{\Lambda}\Gamma^{\Lambda}_{m^{'}n^{'}m n}&=&\int\dfrac{d\omega}{2\pi}\mathcal{S}^{\Lambda}_{k k^{'}}\mathcal{G}^{\Lambda}_{l l^{'}}\left\{
\Gamma^{\Lambda}_{m^{'}n^{'}l k}\Gamma^{\Lambda}_{l^{'}k^{'}m n}\nonumber\right.\\&-&\left[\Gamma^{\Lambda}_{m^{'}k^{'}m l}\Gamma^{\Lambda}_{l^{'}n^{'}k n}+(l\leftrightarrows k,l^{'}\leftrightarrows k^{'})\right]\nonumber\\&+&
\left.\left[\Gamma^{\Lambda}_{n^{'}k^{'}m l}\Gamma^{\Lambda}_{l^{'}m^{'}k n}+(l\leftrightarrows k,l^{'}\leftrightarrows k^{'})\right]\right\},\nonumber
\end{eqnarray}
where each index collects the dot and spin indexes, e.g. $m=(i,\sigma)$ and summation over repeated indexes is assumed,
$\mathcal{G}_{m,m^{'}}^{\Lambda}\equiv\mathcal{G}_{ii',\sigma}^{\Lambda}\delta_{\sigma \sigma'}=\left[\left[\mathcal{G}^{\Lambda}_{0}\right]^{-1}-\Sigma^{\Lambda}\right]_{m m'}^{-1}$ and $
\mathcal{S}^{\Lambda}=\mathcal{G}^{\Lambda}\partial_{\Lambda}\left(\mathcal{G}^{\Lambda}_{0}\right)^{-1}\mathcal{G}^{\Lambda}$ are the dressed Green function and single--scale propagator, respectively, which are matrices in the dots space.

\par
Using the Green function obtained at the end of the fRG flow $\mathcal{G}^{\Lambda\rightarrow 0}(i\omega)$, which provides an approximation for the exact Green function of an interacting quantum dot system, we calculate the average occupation numbers 
\begin{equation}
\langle n_{j,\sigma}\rangle=
\int{\dfrac{d\omega}{2\pi} e^{i\omega 0^{+}}
\mathcal{G}
_{jj,\sigma}^{\Lambda\rightarrow 0}}(i \omega),
\label{occ}
\end{equation}
and the $T=0$ linear conductance $G=\sum_\sigma G_\sigma$, which can be written in the form of Landauer formula as (see, e.g., Ref.~\cite{
Oguri_2001})
\begin{equation}
G_\sigma=2G_{0}\left|\sum_{j,j^{'}}\sqrt{\Gamma_{j}^{R}\Gamma_{j^{'}}^{L}}\mathcal{G}^{\Lambda\rightarrow 0}_{jj^{'},\sigma}(0)\right|^{2},
\label{G(Vg)}
\end{equation}
where $G_{0}={2e^{2}}/{h}$ is the conductance quantum; the Eq.~(\ref{G(Vg)}) can be derived from the Kubo formula, neglecting the vertex corrections, which is justified e.g. for the frequency-independent self-energy~\cite{Karrasch_2006,Enss_thesis,Karrasch_thesis2,Karrasch_thesis}.

\vspace{0.5cm}
\section{fRG results for different types of asymmetry}\par
\subsection{Left-right asymmetry}\par

We first consider the double quantum dots system with up-down symmetry $\Gamma^{L(R)}_{1}=\Gamma^{L(R)}_{2}$, but left-right coupling asymmetry $\Gamma^{R}_{1(2)}=\chi\Gamma^{L}_{1(2)}$, where, without loss of
generality, we assume that $0<\chi\le 1$. In particular, when $\chi=1$, the hopping matrix elements are equal; this case
has been considered within the fRG with counterterm approach previously \cite{IILM}.
In agreement with NRG predictions \cite{Zitko_2007(2)}, in this case
the conductance exhibits a discontinuity at a gate voltage, corresponding to the first order phase transition from SFL to FL phase, and almost reaches the unitary limit value $2e^{2}/h$ at $V_{g}=0$.\par

For an arbitrary parameter $\chi$ a double quantum dot system with the left-right hybridization asymmetry can be effectively considered as a fully hybridization symmetric system with the effective hybridization parameters $\tilde{\Gamma}^{\alpha}_{j}=(1+\chi)\Gamma_{1}^{L}/2$ 
for $\alpha=L,R$ and $j=1,2$. This reflects the fact that the explicit expression for the Green function of the system (see Eq.~(\ref{Green_function})) is invariant under the transformation $\Gamma^{\alpha}_{j}\leftrightarrow \tilde{\Gamma}^{\alpha}_{j}$. In this way, the conductance of of original system $g$ is related to the conductance of the effective system $g_{\rm eff}$ with hybridization parameters $\tilde{\Gamma}^{\alpha}_{j}$ by
\begin{equation}
g(V_{g})=\dfrac{4\chi}{(1+\chi)^{2}}g_{\rm eff}(V_{g}).
\label{G-G_eff}
\end{equation}
Thus, the gate voltage dependence of the conductance for an asymmetric system can be obtained from the one for the symmetric system with dots-leads hybridizations $\tilde{\Gamma}^{\alpha}_{j}<\Gamma^{\alpha}_{j}$,  
by multiplying the latter by a factor $4\chi/(1+\chi)^2<1$, 
which is similar to the Meier-Wingreen formula ~\cite{Meir_1992}. 
This means that left-right coupling asymmetry does not lead to new features in the conductance in respect to those appearing in the symmetric case.

\begin{figure}[t]
\center{\includegraphics[width=1\linewidth]{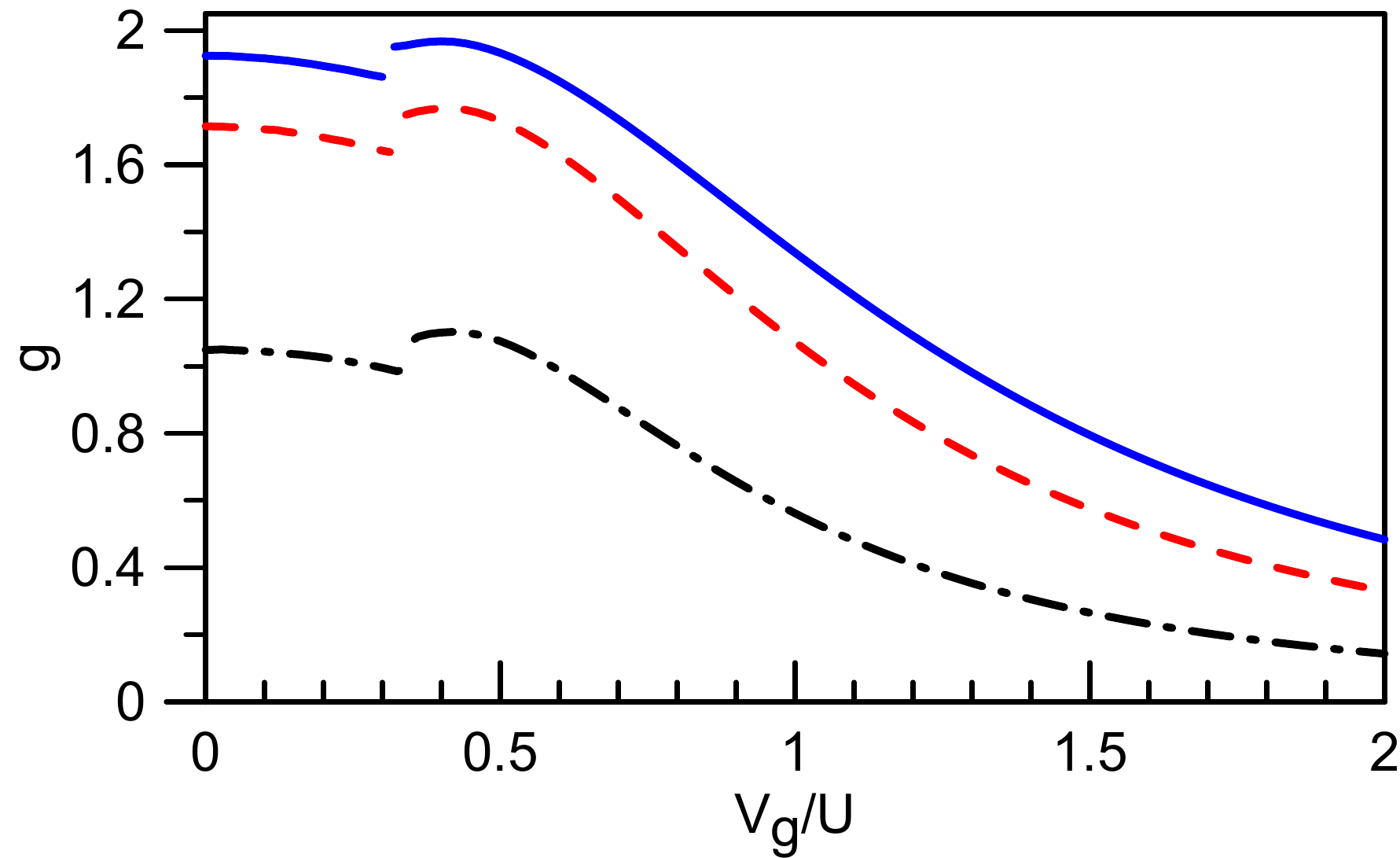}}
\caption{(Color online). The dependence of the linear conductance $g$ on gate voltage ${V_{g}}$ at zero magnetic field $H\rightarrow 0$ and $T=0$ for parallel double quantum dot system with left-right coupling asymmetry: $\Gamma^L_1=\Gamma^L_2=U/4$ and $\Gamma^R_1=\Gamma^R_2=\chi\Gamma^L_1$, with $\chi=0.8$ (blue solid line), 0.5 (red dashed line) and 0.2 (black dashed-dotted line) within the fRG approach with the linear counterterm $(\tilde{H}/U=0.1, \Lambda_{c}/U=0.05)$.
}
\label{left-right}
\end{figure}

In Fig.~\ref{left-right}, the gate voltage dependence of the total conductance $g(V_{g})=2G/G_0$ is shown for 
different left-right asymmetries $\chi=0.8$, $0.5$, $0.2$, and $\Gamma^L_1=U/4$. One can see that, as expected, the behavior of the conductance is quite similar
to the isotropic case: for any choice of the parameter $\chi$, the conductance 	
shows discontinuity caused by a quantum phase transition at a gate voltage $V_{g}=V^{c}_{g}(\chi)$, which weakly depends on the strength of the left-right asymmetry.

As can be seen from Fig.~\ref{left-right}, 
for different parameters $\chi$ the conductance  almost reaches a maximum value of $g^{\rm max}
={8\chi}/{(1+\chi)^{2}}$ at the half-filling $(V_{g}=0)$. The first order phase transition point $V_{g}^{c}(\chi)$ slightly shifts towards higher gate voltages with decreasing $\chi$, which can be attributed to the fact that with decreasing of $\chi$ the
ratio $U/\tilde{\Gamma}_{j}^{\alpha}=2(1+\chi)^{-1}U/\Gamma_{1}^{L}$ 
increases, and consequently, according to the phase diagram in Ref.~\cite{Zitko_2012}, obtained for isotropic quantum dot system within the NRG approach, the region of existence of the SFL phase gradually grows.\par

\subsection{Up-down asymmetry}

\begin{figure}[b]
\center{\includegraphics[width=1\linewidth]{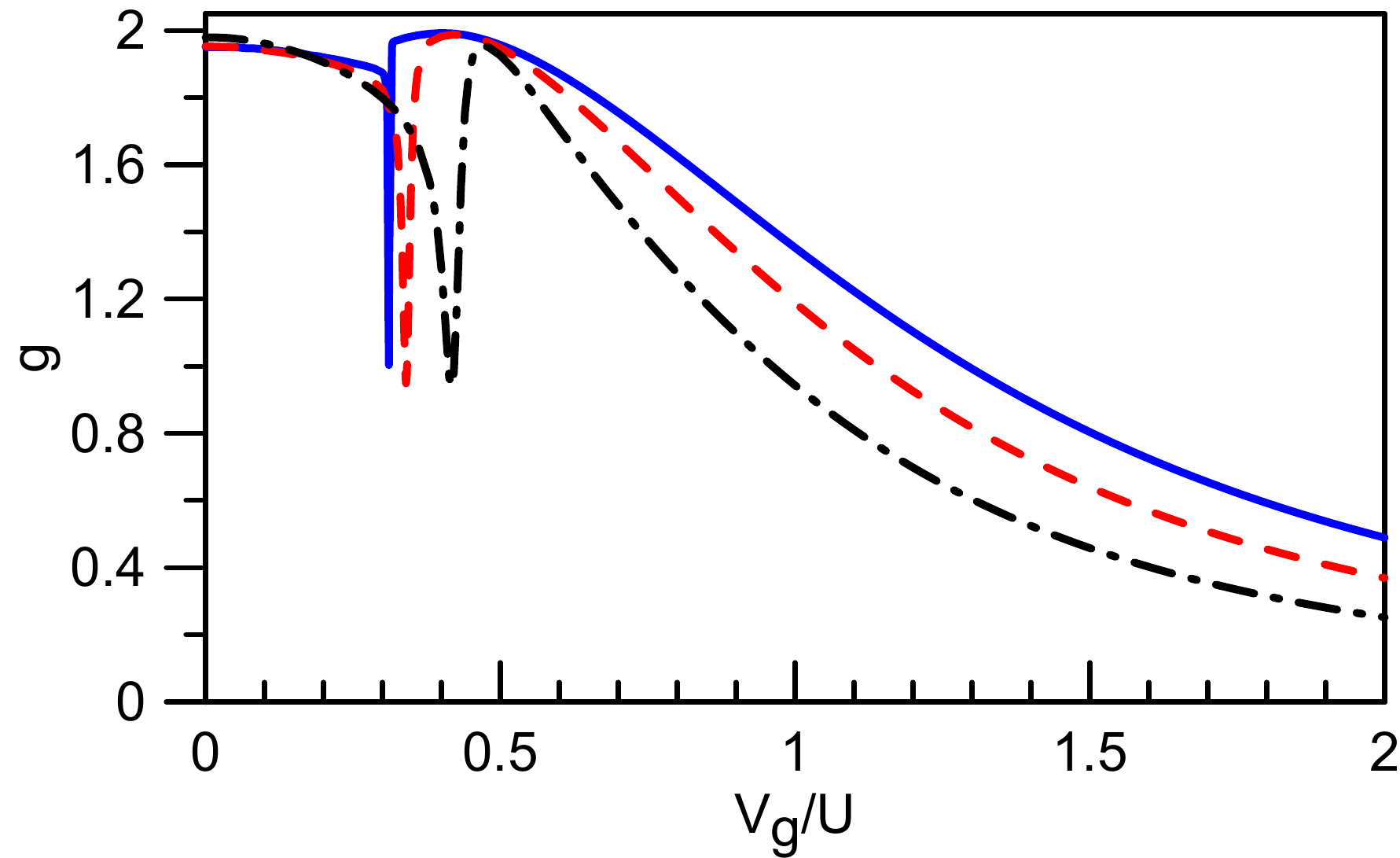}}
\caption{(Color online). The dependence of the linear conductance $g$ on gate voltage ${V_{g}}$  at zero magnetic field $H\rightarrow 0$ and $T=0$ for parallel double quantum dot system with up-down coupling asymmetry: $\Gamma^L_1=\Gamma^R_1=U/4$ and $\Gamma^L_2=\Gamma^R_2=\gamma \Gamma^L_1$, with $\gamma=0.8$ (blue solid line), 0.5 (red dashed line) and 0.2 (black dashed-dotted line) within the fRG approach with the linear counterterm $(\tilde{H}/U=0.1, \Lambda_{c}/U=0.05)$.
}
\label{up-down}
\end{figure}
In case of up-down hybridization asymmetry $\Gamma^{L(R)}_{2}=\gamma\Gamma^{L(R)}_{1}$, the gate voltage dependence of the conductance changes due to the generation of the effective electron hopping between even and odd orbitals 
during the fRG flow, as discussed below.
In Fig.~\ref{up-down} the dimensionless conductance $g$ as a function of the gate voltage $V_{g}$ for $\Gamma^{L(R)}_{1}=U/4$ is plotted for several values of the asymmetry parameter $\gamma=0.2$, $0.5$, $0.8$. One can see that for different values of the parameter $\gamma$ the conductance $g(V_g)$ is continuous and exhibits the distinct sharp asymmetric anti-resonance at some gate voltage, 
depending on the parameter $\gamma$; when $\gamma$ increases, the anti-resonance becomes narrower and its position shifts to lower gate voltages.
At gate voltages $|V_{g}|$ smaller the position of the anti-resonance, the conductance  increases with decreasing $|V_{g}|$ and near half--filling $(V_{g}=0)$ it  
almost reaches the unitary limit value $g(0)=2$ $(G=2e^{2}/{h})$.  As discussed in Ref.~\cite{IILM}, this 
behavior of the conductance is an indication of a singular Fermi-liquid (local moment) state at sufficiently small gate voltages.

To explain the observed features of the conductance, we pass to the even-odd orbitals (see explicit form of the even-odd transformation in Appendix), and rewrite the conductance per spin in fRG approach $g_{\sigma}=2G_{\sigma}/G_0$ in the form
\begin{equation}
g_{\sigma}=\dfrac{4\Gamma^{L}_{e}\Gamma^{R}_{e}}{q^{2}_{\sigma}+\Gamma_{e}^{2}},
\label{g_sigma}
\end{equation}
where $\Gamma^{\alpha}_{e}=\pi|t^{\alpha}_{e}|^{2}\rho_{\rm lead}(0)$ and $t^{\alpha}_{e}=\sqrt{1+\gamma}t^{\alpha}_{1}$ are the hybridization and hopping from the leads to the even state, 
$\Gamma_{e}=\Gamma^{L}_{e}+\Gamma^{R}_{e}$ is the total level broadening of the even state. The parameter $q_{\sigma}=(\epsilon_{o,\sigma}\epsilon_{e,\sigma}-(t_{eo}^{\sigma})^{2})/\epsilon_{o,\sigma}$ is determined by the renormalized energies of even and odd states
\begin{equation}
\epsilon_{e(o),\sigma}=\left[\epsilon_{1(2),\sigma}+\eta^{2}\epsilon_{2(1),\sigma}\mp 2\eta t^{\sigma}_{12}\right]/(1+\eta^{2}),
\label{eeo}
\end{equation}
and the effective (renormalized) hopping parameter between even and odd orbitals 
\begin{equation}
t^{\sigma}_{eo}=\left[\eta\left(\epsilon_{1,\sigma}-\epsilon_{2,\sigma}\right)-t^{\sigma}_{12}\left(\eta^{2}-1\right)\right]/\left(1+\eta^{2}\right),
\label{Teo}
\end{equation}
$\epsilon_{j,\sigma}=\epsilon_{\sigma}+\Sigma^{\Lambda\rightarrow 0}_{jj,\sigma}$ and $t^{\sigma}_{ij}=-\Sigma^{\Lambda\rightarrow 0}_{ij,\sigma}$ correspond to the renormalized quantum dot energy levels and interdot hopping parameters, respectively, $\eta=\gamma^{1/2}$. As it is shown in the Appendix, in the absence of magnetic field the total conductance can be obtained from that, containing only one spin projection: $g=g_{\uparrow}+g_{\downarrow}=g_{\sigma}(V_{g})+g_{\sigma}(-V_{g})$. 
\par
In the absence of the up-down asymmetry $\gamma=1$ (considered in previous subsection) the conductance depends only on the position of the renormalized even energy level,  
$g_{\sigma}=({4\Gamma^{L}_{e}\Gamma^{R}_{e}})/({\epsilon_{e,\sigma}^{2}+\Gamma_{e}^{2}})$, and does not vanish for any gate voltage, reaching maximal value $g^{\rm max}_{\sigma}=({4\Gamma^{L}_{e}\Gamma^{R}_{e}})/\Gamma_{e}^{2}$.
One can see from Eq.~(\ref{Teo}), however, that for $\gamma\ne 1$ the renormalized hopping parameter $t^{\sigma}_{eo}$ is not zero identically and hence the conductance also 
depends on the position of 
the odd orbital energy level. 
This level has a non-trivial effect on the conductance, in particular 
complete suppression of $g_\sigma$
is possible
when $\epsilon_{o,\sigma}=0$ due to destructive interference between the contributions of even and odd states. On the other hand, the maximum value of the conductance $g^{\rm max}_{\sigma}$ 
is realized if the energy levels $\epsilon_{e/o,\sigma}$ and hopping parameter $t_{eo}^{\sigma}$ fulfill the relation $q_\sigma=0$.

\begin{figure}[h]
\center{\includegraphics[width=0.95\linewidth]{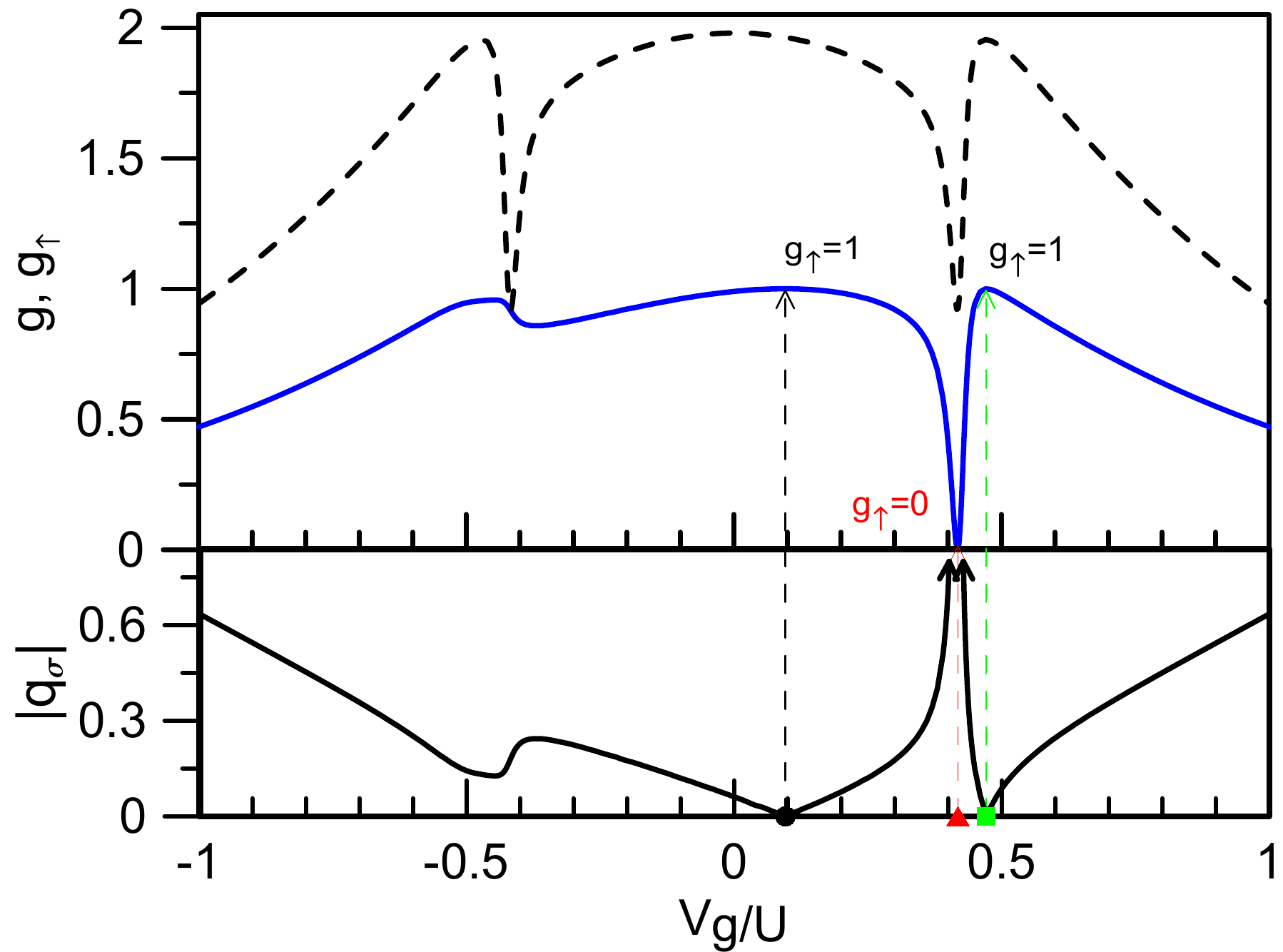}} 
\caption{(Color online). Upper panel: gate voltage dependencies of the total conductance $g=g_{\uparrow}+g_{\downarrow}$ (black dashed line) and spin-up conductance $g_{\uparrow}$ (blue solid line).  Lower panel: gate voltage dependence of the parameter $|q_{\uparrow}|$ (see text). The parameters are the same as in the case of $\gamma=0.2$ of Fig.~\ref{up-down}.}
\label{up-down_0.2}
\end{figure}

\begin{figure}[h]
\center{\includegraphics[width=0.95\linewidth]{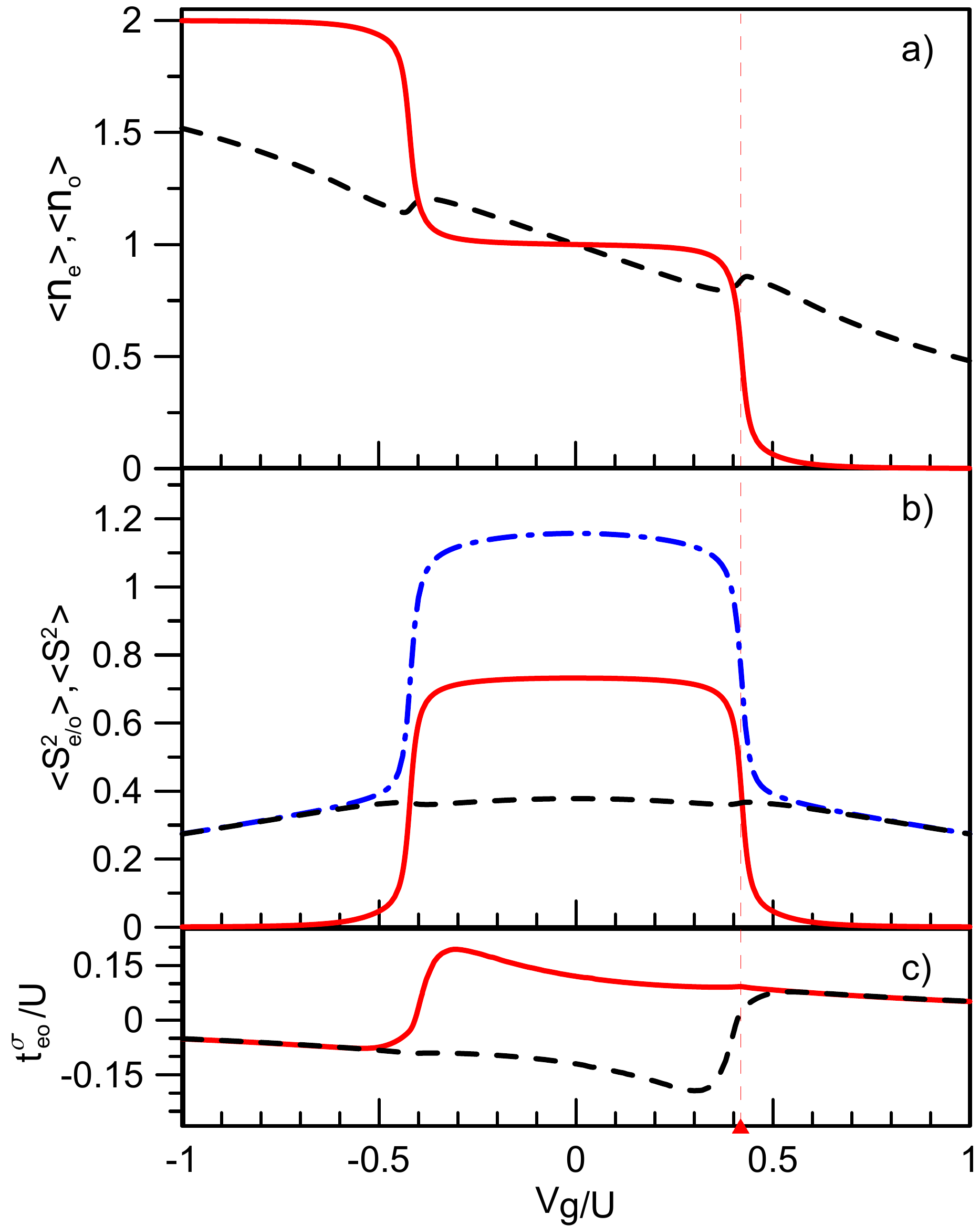}}
\caption{(Color online). The average occupation numbers of the odd $\langle n_{o}\rangle$ (red solid line) and even  $\langle n_{e}\rangle$ (black dashed line) states (a), the average square of magnetic moment $\langle\mathbf{S}_{e(0)}^{2}\rangle$ in the odd (red solid line) and even (black dashed line) states, as well as the average of the square of the total spin $\langle\mathbf{S}_{t}^{2}\rangle=\langle\left(\mathbf{S_{1}+S_{2}}\right)^{2}\rangle$ (blue dashed-dotted line) (b), and the hopping parameter between the even and odd levels (c) $t^{\sigma}_{eo}$ (solid red/dashed black line for $\sigma=\uparrow/\downarrow$) as a  function of the gate voltage $V_{g}$ for $\gamma=0.2$ (the other parameters are  the same as in Fig.~\ref{up-down}). The vertical dashed line correspond to gate voltage $V_g^{(2)}$ (red triangle), 
introduced in Fig. \ref{up-down_0.2} and discussed in the text.} 
\label{neo-seo}
\end{figure}

In Fig.~\ref{up-down_0.2} the $V_{g}$ dependence of the conductance $g_{\uparrow}$ and the absolute value of the parameter $q_{\uparrow}$ for the spin-up electrons are plotted for one of the parameter sets of Fig. \ref{up-down}, $\Gamma^L_1=\Gamma^R_1=U/4$ and $\gamma=0.2$. One can see that $q_{\uparrow}$ becomes zero at the
value of the gate voltage $V_{g}=V^{(1)}_{g}$ (marked by black filled circle), which is close to half-filling $V_{g}=0$, and at the gate voltage $V_{g}=V^{(3)}_{g}$ (green filled square). From previous consideration it follows that both these values of the gate voltages yield the conductance maximum (see Eq.~(\ref{g_sigma}))  $g_{\uparrow}=g^{\rm max}_{\uparrow}={4\Gamma^{L}_{e}\Gamma^{R}_{e}}/{\Gamma_{e}^{2}}=1$  (see Fig.~\ref{up-down_0.2}a). For $V_{g}<0$, $q_{\uparrow}$ behaves smoothly and its absolute value reaches a minimum near the gate voltage $V_{g}=-V_{g}^{(3)}$, which results in the maximum of the spin-up/spin-down conductance 
in the vicinity of $V_{g}=\mp V_{g}^{(3)}$. Consequently, the total conductance $g(V_{g})=g_{\uparrow}(V_{g})+g_{\uparrow}(-V_{g})
$ exhibits maximum at $V_{g}=V_{g}^{(3)}$ with $g(V_{g}^{(3)})\approx 2$. The same value of the conductance is obtained close to half filling due to maximum of $g_\uparrow$ at $V_g=V_g^{(1)}$.

On the other hand, the parameter $q_{\uparrow}$ diverges when $V_{g}\rightarrow V^{(2)}_{g}$  (red filled triangle) because of the crossing of the odd energy level $\epsilon_{o,\sigma}$ the Fermi level of the lead, which is put to zero.
Consequently, according to Eq.~(\ref{g_sigma}), conductance $g_\uparrow$ abruptly falls, 
vanishing at  $V_{g}=V^{(2)}_{g}$. This corresponds to the above-mentioned sharp anti-resonance with $g(V_{g}^{(2)})\approx 1$.

  To study the relation of the observed features of the conductance to the formation of local moments, we consider occupation numbers and the square of the spin. By using again the transformation to the even-odd orbitals, the total average occupation numbers for each spin
direction $\langle n_{\sigma}\rangle=\sum_{j}{\langle n_{j,\sigma}\rangle}$ can be written explicitly as
$$
\langle n_{\sigma}\rangle=1-\dfrac{1}{\pi}\arctan{
\dfrac{q_\sigma}{\Gamma_{e}}
}
-\dfrac{1}{2}\sgn{\left(\epsilon^{\sigma}_{o}\right)},
$$
and thus each minimum $g_{\sigma}=0$ or maximum $g_{\sigma}=1$ value of the partial conductance corresponds to integer $\langle n_{\sigma}\rangle\in\{0,1,2\}$  or half-integer $\langle n_{\sigma}\rangle\in\{1/2,3/2\}$ values of the occupation numbers, respectively.\par
In Fig. \ref{neo-seo} we plot the occupation numbers $\langle n_{e(o)}\rangle$ and the square of the spin
$\langle \mathbf{S}_{e/o}^2 \rangle$, corresponding to the even and odd orbitals (see  Appendix) for strong anisotropy of hopping parameters $\gamma=0.2$. One can see that at small gate voltages $V_g$ (i.e. close to half filling) $\langle n_o\rangle \simeq 1$ and there is substantial square of the local moment $\langle {\bf S}_o^2\rangle \simeq 3/4$. Both, $\langle n_o\rangle$ and $\langle {\bf S}_o^2\rangle$  change continuously, dropping sharply at the critical gate voltage $V_g^{c}$, 
coinciding with the above introduced gate voltage $V_g^{(2)}$, at which the conductance reaches minimum. The continuous change of these parameters is due to generation of the effective hopping between the odd and even orbitals (see Fig. \ref{neo-seo}c).
The gate voltage $V_g^c=V_g^{(2)}$ can be therefore identified with the quantum phase transition point from the singular to the regular Fermi-liquid phase. We have verified by performing additional  numerical renormalization-group calculations, that small finite value of $\langle S_o^2\rangle$ (related to small spin splitting of energy levels) at $|V_g|>V_g^c$, 
is an artifact of the considered fRG method, but the transition remains continuous. Apart from the narrow vicinity of the transition (where qualitatively correct results are obtained at $|V_g|<V_g^c$), the considered approach describes the behavior of conductance and occupation numbers quantitatively correct.

One can see therefore, that the local moment in the odd orbital is almost fully preserved even for rather strong up-down asymmetry. The reason is that, as well as for a perfectly symmetric case ~\cite{IILM}, the spin splitting of the energy levels in infinitesimally small magnetic field is provided by the "Hund" term in the Hamiltonian, rewritten in terms of even and odd states (see Appendix), which appears to be of the order of the interaction strength $U$. At the same time, the generated hopping between even and odd orbitals is much smaller, $|t^{\sigma}_{eo}|\ll U$, see Fig. \ref{neo-seo}c, and therefore it does not destroy the local moment in the odd state even for rather strong asymmetry. 

\begin{figure}[t]
\center{\includegraphics[width=1\linewidth]{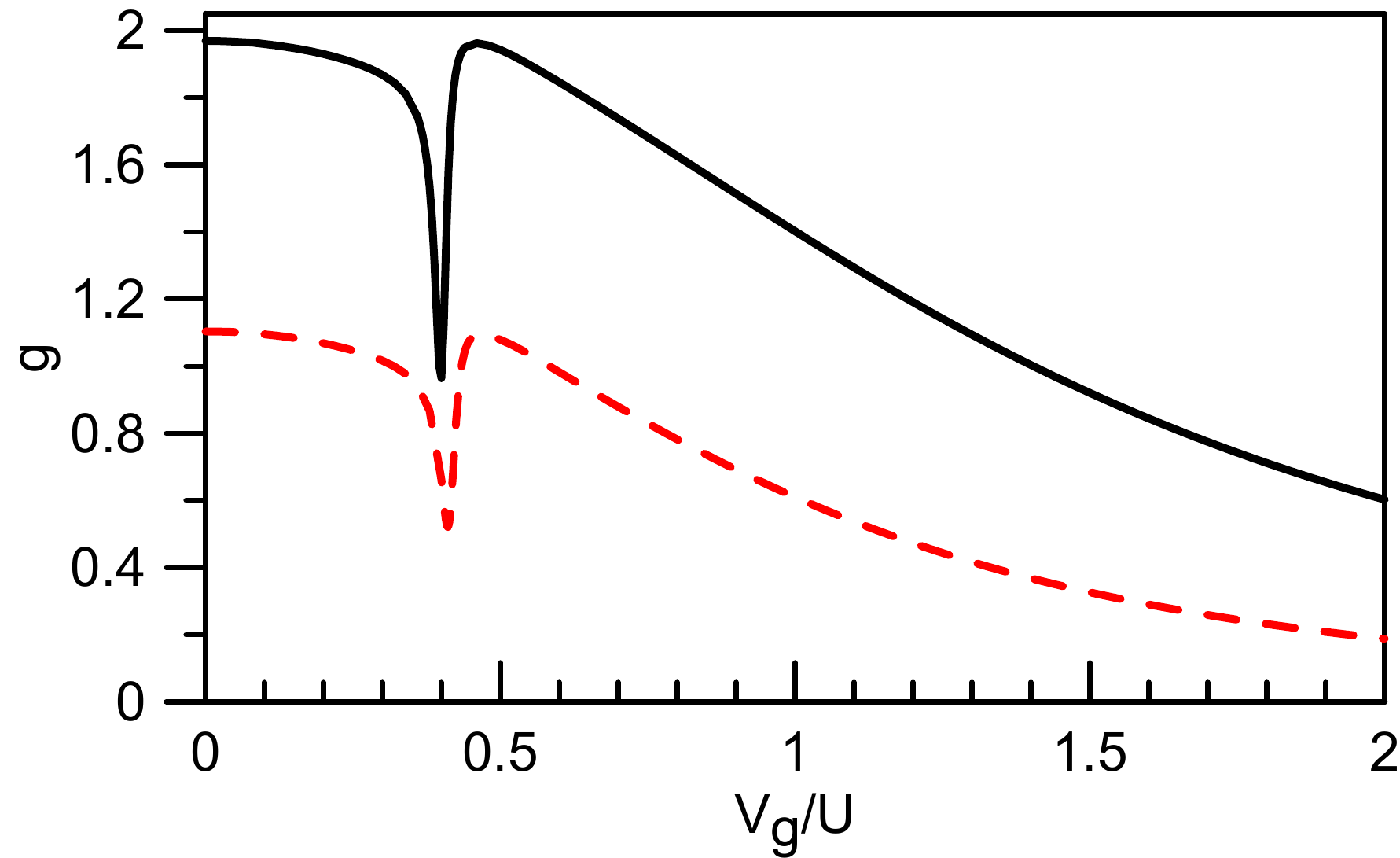}}
\caption{(Color online). The dependence of the linear conductance $g$ on gate voltage ${V_{g}}$ at zero magnetic field $H\rightarrow 0$ and $T=0$ for parallel double quantum dot system with mixed coupling asymmetry: $\Gamma^{L(R)}_{2}=\gamma\Gamma^{L(R)}_{1}$, $\Gamma^{R}_{1(2)}=\chi\Gamma^{L}_{1(2)}$ with $(\gamma, \chi)=(0.2,0.8)$ (solid black line), (0.2,0.2) (red dashed line) and $\Gamma^L_1=U/2$ within the fRG approach with the  counterterm $(\tilde{H}/U=0.1, \Lambda_{c}/U=0.05)$.
\label{mix}
}
\end{figure}
\subsection{Mixed asymmetry}\par
In case of
both, up-down and left-right types of asymmetry, $\Gamma^{L(R)}_{2}=\gamma \Gamma^{L(R)}_{1}$, $\Gamma^{R}_{1(2)}=\chi\Gamma^{L}_{1(2)}$, $0<\gamma<1$, $0<\chi<1$, analogously to the previous consideration, instead
of the initial quantum dot system one can consider the effective system, which has only the up-down asymmetry with $\tilde{\Gamma}^{L(R)}_{1}=(1+\chi)\Gamma_{1}^{L}/2$ and $\tilde{\Gamma}^{L(R)}_{2}=\gamma(1+\chi)\Gamma_{1}^{L}/2$.  
Then, the expression for the conductance can be written in the form Eq.~(\ref{G-G_eff}), where $g_{\rm eff}(V_{g})$ represents now the gate voltage dependence of the conductance for an effective quantum dot system with up-down coupling asymmetry. Therefore, for fixed Coulomb interaction both the conductance curve differ only by a constant factor, which depends on the left-right asymmetry of the system. As an example, in Fig.~\ref{mix} we plot the gate voltage dependence of the linear conductance for $U=2\Gamma_{1}^{L}$ and the following configurations of asymmetry: $(\gamma,\chi)=(0.2,0.2)$, $(0.2,0.8)$. As expected the conductance behaves the same way as in the up-down asymmetry case and as in the previous cases for $V_{g}=0$ shows the  maximum value $g^{\rm max}=8\chi/(1+\chi)^{2}<2$, which does not depend on the up-down asymmetry parameter $\gamma$.


\subsection{General asymmetry}
In more complicated cases, when the dots-leads hopping parameters are fully independent, 
one can use the transformation to some effective even and odd states, which are chosen according to some criterion. In general, the coupling of the effective odd orbital to the leads does not vanish, and there is no fully local moment in the odd state even at $V_g=0$; however, as we will see below, the local moment can be almost formed in the sense that $\langle {\bf S}_o^2\rangle \simeq 3/4$.  \par
In the presence of general asymmetry, the effective ``even" and ``odd" energy levels can be determined similarly to previous sections and are given by (see Appendix)
$$
\epsilon_{e(o),\sigma}=a^{2}\epsilon_{1(2),\sigma}+(1-a^{2})\epsilon_{2(1),\sigma}\mp 2a(1-a^{2})^{1/2}t_{12}^{\sigma},
$$
while the effective ``even"--``odd" state hopping parameter is
$$
t_{eo}^{\sigma}=a(1-a^{2})^{1/2}\left(\epsilon_{1,\sigma}-\epsilon_{2,\sigma}\right)+(2 a^{2}-1)t_{12}^{\sigma},
$$
where the parameter $a$ is related to previously used in Eqs. (\ref{eeo}) and (\ref{Teo}) parameter $\eta$ by $a=(1+\eta^2)^{-1/2}$, but 
its relation to asymmetry of the hybridizations is more involved.
Specifically, we will determine the parameter $a$ from the condition of the minimum of the coupling between leads and the ``odd" orbital,
\begin{equation}
F(a)=|t^{L}_{o}(a)|+|t^{R}_{o}(a)|,
\label{cond1}
\end{equation}
where $t^{\alpha}_{o}(a)=a t^{\alpha}_{2}-(1-a^{2})^{1/2}t^{\alpha}_{1}$.
This way we find
\begin{equation}
a=\begin{cases}
t^{L}_{1}/t^{L},&\text{  }t^{L}\ge t^{R};\\
t^{R}_{1}/t^{R},&\text{  }t^{L}\le t^{R},
\end{cases}
\label{a_tr}
\end{equation}
where $t^{\alpha}=\sqrt{\left(t_{1}^{\alpha}\right)^{2}+\left(t_{2}^{\alpha}\right)^{2}}$.\par
\begin{figure}[t]
\center{\includegraphics[width=0.95\linewidth]{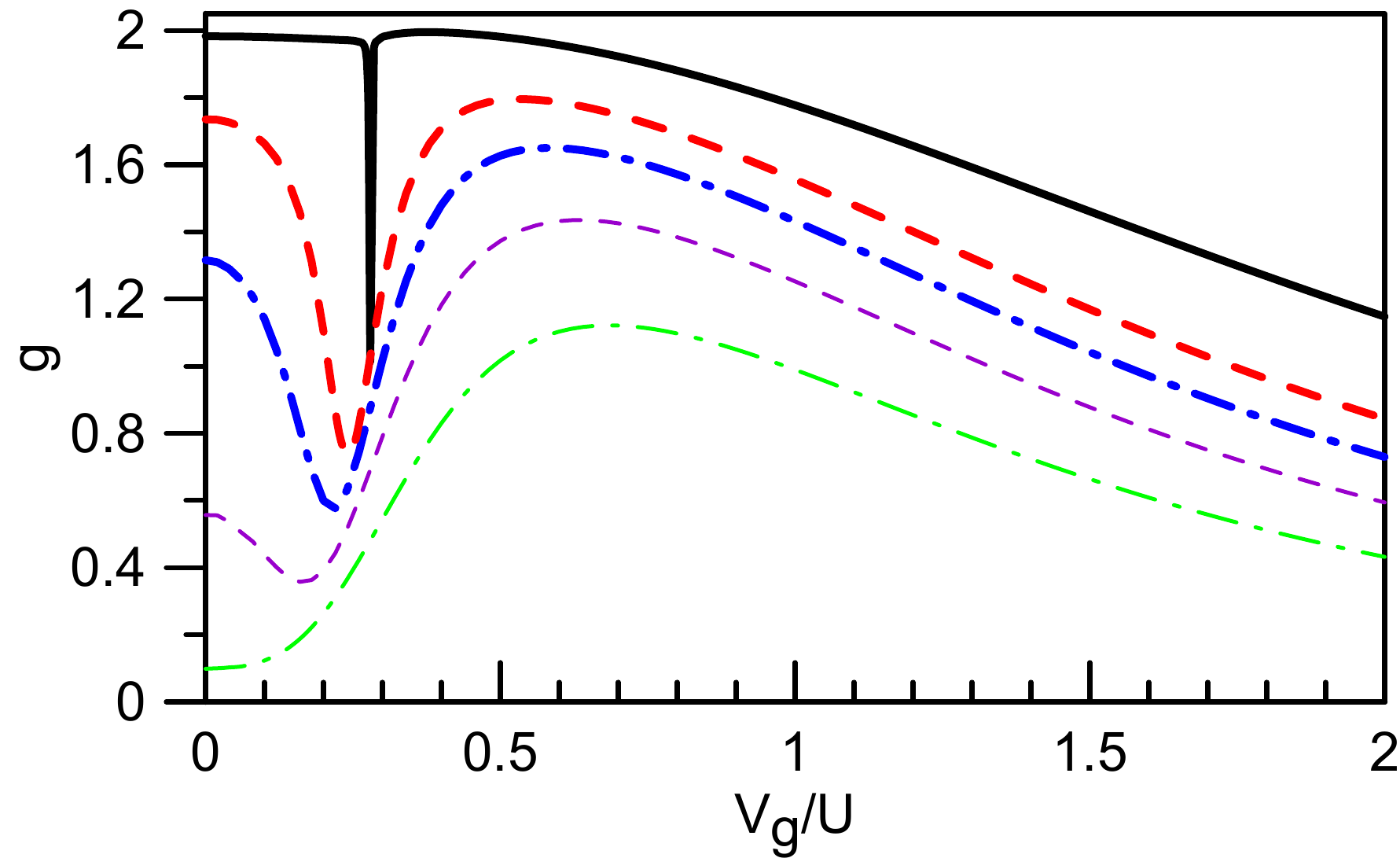}}
\caption{(Color online). The dependence of the linear conductance $g$ on gate voltage ${V_{g}}$ at zero magnetic field $H\rightarrow 0$ and $T=0$ for parallel double quantum dot system with diagonal symmetric coupling: $\Gamma^L_1=\Gamma^R_2=U/2$, $\Gamma^L_2=\Gamma^R_1=\Gamma^{RL}_{12}$. From upper to lower curve $\Gamma^{RL}_{12}/U=0.45, 0.25, 0.2, 0.15, 0.1$, respectively.  The calculations were performed within the fRG approach with the counterterm $(\tilde{H}/U=0.1, \Lambda_{c}/U=0.05)$.   
} 
\label{G_diag}
\end{figure}
\subsubsection{Diagonal asymmetry}\par
Let us first consider the case of diagonal coupling asymmetry
$t_{1(2)}^{L}=t_{2(1)}^{R}$ $(\Gamma_{1(2)}^{L}=\Gamma_{2(1)}^{R})$, in which, except for case of $t_{1}^{L}=t_{1}^{R}$ (when the system is completely symmetric), $t_{1}^{L}/t_{2}^{L}\ne t_{1}^{R}/t_{2}^{R}$. In Fig.~\ref{G_diag} the gate voltage dependence of the conductance is shown for $\Gamma_{1}^{L}=\Gamma_{2}^{R}=U/2$ and different values of $\Gamma_{1}^{R}=\Gamma_{2}^{L}=\Gamma^{RL}_{12}$. One can see that the behavior of the conductance strongly depends on the system asymmetry. For not too strong deviation from the isotropic case $(\Gamma^{RL}_{12}=0.45)$ the conductance (thick solid line) behaves similarly to the up-down asymmetry case, showing a sharp asymmetric anti-resonance. With increasing asymmetry of the system, the gate voltage dependence of the
conductance changes significantly and value of the conductance at $V_{g}=0$ decreases. For intermediate asymmetries $(\Gamma^{RL}_{12}=0.25)$ (thick dashed line) and $(\Gamma^{RL}_{12}=0.2)$ (thick dashed-dotted line), the 
anti-resonance is preserved, but its width becomes larger as  $\Gamma^{RL}_{12}$ decreases. For sufficiently large asymmetry we find that the  above-discussed form of the resonance disappears and the conductance is strongly suppressed near zero gate voltage. \par
\begin{figure}[t]
\center{\includegraphics[width=0.95\linewidth]{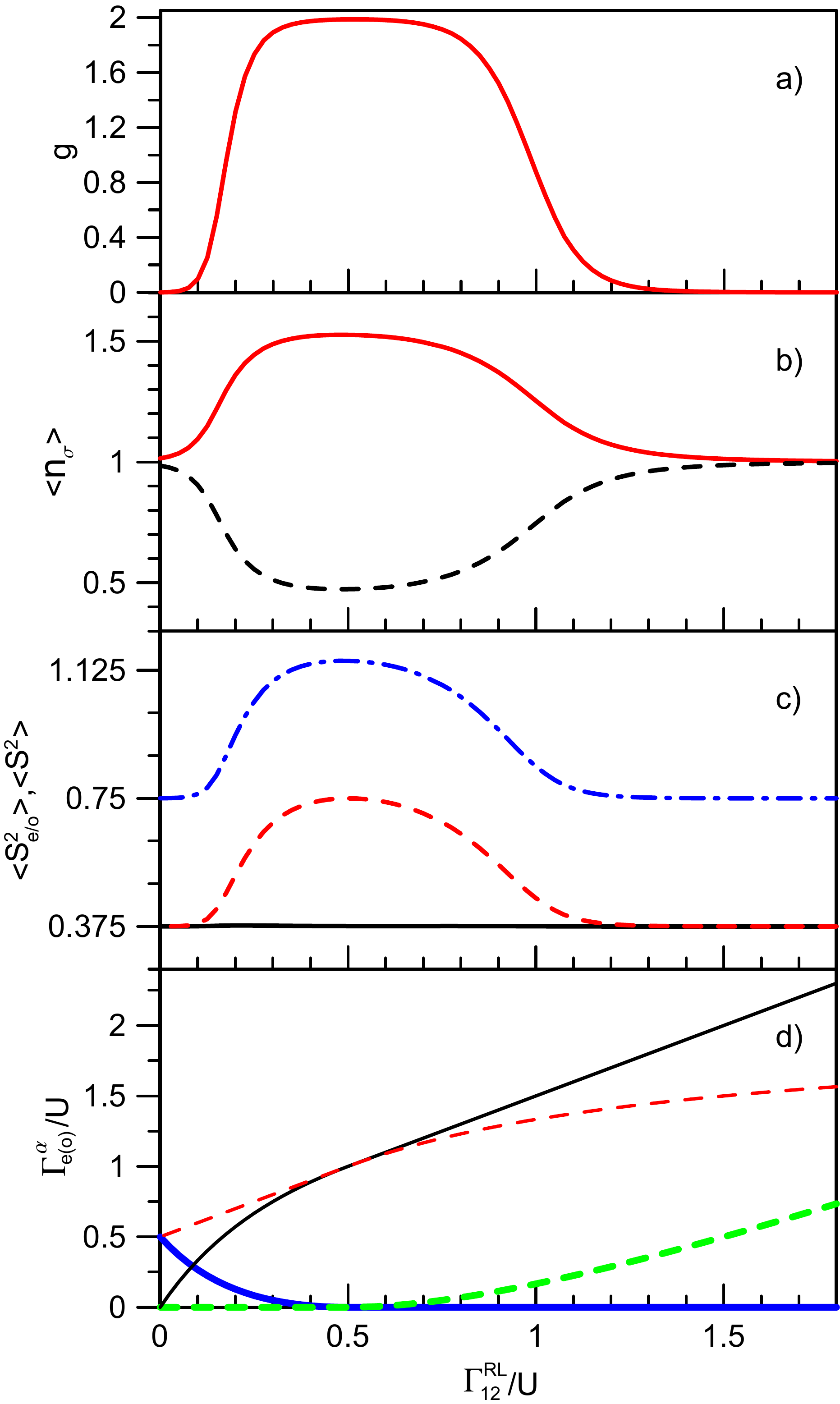}}
\caption{(Color online). The  conductance (a), the average occupation numbers (b) $\langle n_{\sigma}\rangle=\sum_{j}\langle n_{j,\sigma}\rangle$ (solid (red) line for $\sigma=\uparrow$ and dashed (black) line for $\sigma=\downarrow$), the average square of magnetic moment $\langle\mathbf{S}_{e(o)}^{2}\rangle$ in the even (solid (black) line) and odd (dashed (red) line) states, as well as the average of the square of the total spin $\langle\mathbf{S}_{t}^{2}\rangle=\langle\left(\mathbf{S_{1}+S_{2}}\right)^{2}\rangle$ (dashed-dotted (blue) line) (c), and the hybridization between leads and even $\Gamma^{\alpha}_{e}$ (thin solid (black) line for $\alpha=L$ and thin dashed (red) line for $\alpha=R$) and odd $\Gamma^{\alpha}_{o}$ (thick solid (blue) line for $\alpha=L$ and thick dashed (green) line for $\alpha=R$) orbitals (d) as a function of $\Gamma^{RL}_{12}/U$ for parallel double quantum dot system with diagonal symmetric coupling with $\Gamma^{L}_{1}=\Gamma^{R}_{2}=U/2$  and $V_{g}=0$, $H\rightarrow 0$ within the fRG approach with the counterterm $(\tilde{H}/U=0.1, \Lambda_{c}/U=0.05)$.}
\label{G_N_S2_diag}
\end{figure}
From this gate voltage dependence  of the conductance one can guess partial formation of local moments near half filling in a rather broad range of asymmetries $0.25\lesssim\Gamma^{RL}_{12}<0.5$. This is confirmed by plotting the  $\Gamma_{12}^{RL}$ dependence of the conductance at $V_{g}=0$ (see Fig.~\ref{G_N_S2_diag}a), which has an asymmetric bell-shaped form with the maximum $g\approx 2$ at the symmetric point $\Gamma_{12}^{RL}=0.5$.
As can be seen from  Fig.~\ref{G_N_S2_diag}b, large conductance at zero gate voltage corresponds to an essential spin splitting of electronic states in the considering limit $H\rightarrow 0$, which is similar to previously considered isotropic case \cite{IILM} and above discussed cases of left-right and up-down asymmetries.
To show explicitly that the above-considered behavior of the conductance is closely related to  the presence of the
partially formed local magnetic moment on the quantum dots, we plot the square of the moment at the ``even" and ``odd" orbitals (see Fig.~\ref{G_N_S2_diag}c), introduced according to the recipe, outlined above. One can see, that the moment on the ``odd" orbital is peaked in the same range of asymmetries as the conductance and spin splitting. From Fig.~\ref{G_N_S2_diag}d one can see that the hybridization of the ``odd" orbital with the leads in the respective asymmetry range is sufficiently small, which provides a possibility of the existence of local moment. 
\vspace{-0.5cm}
\subsubsection{Arbitrary asymmetry}
We have assumed so far that the hybridization parameters are linked to each other by some relations, which leads to a certain symmetry of the system.  For completeness, in this section, as an example, we consider the quantum dot system, in which all hybridization parameters are independent.\par
\begin{figure}[t]
\center{\includegraphics[width=0.9\linewidth]{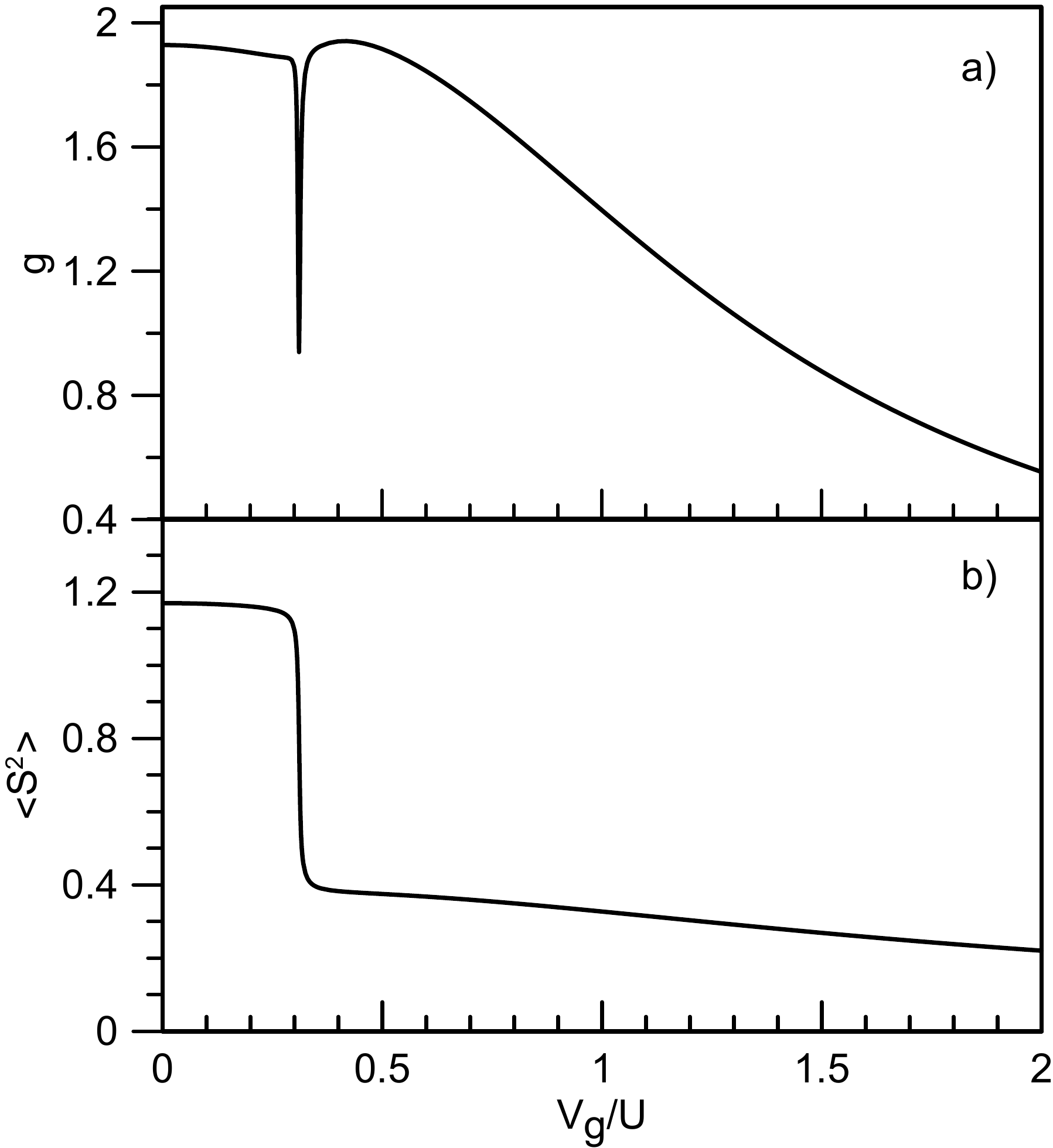}}
\caption{(Color online). The conductance (a) and average of the square of the total spin (b) as a function of the gate voltage $V_{g}/U$ for parallel double quantum dot system with $\Gamma^{L}_{1}/U=0.27$, $\Gamma^{L}_{2}/U=0.16$, $\Gamma^{R}_{1}/U=0.33$, $\Gamma^{R}_{2}/U=0.24$ and $H\rightarrow 0$ within the fRG approach with the counterterm $(\tilde{H}/U=0.1, \Lambda_{c}/U=0.05)$.}
\label{G_S2_arbitrary}
\end{figure}

The conductance and average of the square of the total spin for system with  $\Gamma^{L}_{1}/U=0.27$, $\Gamma^{L}_{2}/U=0.16$, $\Gamma^{R}_{1}/U=0.33$, $\Gamma^{R}_{2}/U=0.24$ are plotted as functions of the gate voltage in Fig.~\ref{G_S2_arbitrary} (for the purpose of comparison with previous results we use the same ratio of hybridizations with the leads as in Ref.~\cite{Karrasch_2006}, but with somewhat smaller interaction strength $U$). The conductance (see Fig.~\ref{G_S2_arbitrary}a) shows overall feature, observed in Sect. IIIB for the up-down asymmetry - the presence of the anti-resonance, accompanied by increase of conductance at small gate voltage, which is characteristic for the partial local moment formation.
From Fig.~\ref{G_S2_arbitrary}b we can conclude that even for this rather general hybridization parameters the local moment can be rather well defined near the half-filling. 
The formation of a local moment near the  half-filling for the present set of parameters can be easily understood on the basis of ``even-odd" states, defined in the beginning of Section IIID. Indeed, by 
passing to these states, the 
system can be mapped onto one  with hybridizations $\Gamma^{L}_{e}\approx0.429$, $\Gamma^{R}_{e}\approx0.570$, $\Gamma^{L}_{o}\approx0.001$ and $\Gamma^{R}_{o}\approx0$, which means that the ``odd" orbitals are almost disconnected from the leads. 
Thus, the condition $\Gamma^{\alpha}_{o}\ll\Gamma^{\alpha}_{e}$ can be viewed as a general criterion of the presence of the partially formed local magnetic moment on quantum dots in the presence of arbitrary asymmetry.\par
\section{Conclusion}
In the present paper, within the counterterm extension of the fRG approach \cite{IILM}, we have performed a systematic study of the effects of asymmetric coupling between the dots and leads on the conductance and possibility of a local magnetic moment formation, for a parallel double quantum dot system.\par
First, we have examined the quantum dot systems, in which one (odd) orbital can be completely disconnected from the leads by an appropriate canonical transformation to the even-odd basis. 
This concerns the case of up-down $\Gamma^{L(R)}_{2}=\gamma\Gamma^{L(R)}_{1}$ and left-right $\Gamma^{R}_{1(2)}=\chi\Gamma^{L}_{1(2)}$, $0<\gamma ,\chi< 1$, types of asymmetry.
In this case the quantum dot system can be viewed as the effective system with only  up-down asymmetry, with the same asymmetry parameter $\gamma$. At the same time, the conductance for these systems differ only by the constant factor, depending on the left-right asymmetry parameter $\chi$ and consequently this mixed type of asymmetry inherits the typical behavior of the conductance for the case of up-down asymmetry.\par

In particular, for up-down symmetry $\Gamma^{L(R)}_{2}=\Gamma^{L(R)}_{1}$ (Sect. IIIA of the paper), the system, like for the isotropic case \cite{Zitko_2007(2),IILM}, shows the first order quantum phase transition to SFL phase accompanied by the jump-like (discontinuity) change of the conductance at the transition point for any choice of the left-right asymmetry $0<\chi< 1$. 
In the case, when up-down symmetry is absent ($\gamma\ne 1$) (Sect. IIIB,C of the paper), a well-defined local magnetic moment also occurs even for rather strong up-down asymmetry (small $\gamma$), which is confirmed by the substantial increase of the square of the local moment for the odd state near the half-filling almost up to the value $\langle {\bf S}_o^2\rangle=3/4$. In contrast to the up-down symmetry case, the conductance is continuous and exhibits sharp asymmetric anti-resonance at the transition point to the SFL state. We have found that the appearance of the anti-resonance is related to the contribution of the odd state, which for the case of up-down asymmetry provides suppression of the conductance when the energy of the odd state coincides with the Fermi level of the leads. With decreasing the up-down asymmetry (increasing the parameter $\gamma$) the anti-resonance in the conductance becomes narrower and its position shifts towards lower gate voltages. \par
We have also considered quantum dot systems with more general asymmetry.
By constructing of the effective ``even-odd" states from the requirement of the minimal (although in general non-zero) absolute value of coupling of the ``odd" orbital to the leads we have shown that the almost formed local moment can occur in a broad range of asymmetries, for which the hybridization of the effective ”odd” orbitals is sufficiently small. In particular, we have demonstrated that the partial local moment formation takes place for not too strong diagonal coupling asymmetry $\Gamma_{1(2)}^{L}=\Gamma_{2(1)}^{R}$, $\Gamma_{1}^{L}\neq\Gamma_{2}^{L}$, and also for a particular example of parallel quantum dot system with rather different hybridization parameters.

Although this paper has focused on the parallel double quantum dot system, we expect similar behavior in other ring geometry systems with larger number of quantum dots, which, however, require further investigations. In particular, used in the present study fRG apprach with the counterterm can be extended to consideration of more complicated quantum dot systems and non-equilibrium situations.  
The possibility to manipulate  the formation of local moment and conductance by small changes of gate voltage even in the presence of moderate asymmetry of hybridizations  to the leads, studied in the present paper, may be useful in nanoscopic devices.  In this respect, more realistic multi-level quantum dot systems and contacts with realistic density of states also require further consideration.

{\it Acknowledgements.} The work is performed within the theme Electron 01201463326 of FASO, Russian Federation and partly supported by RFBR grant 17-02-00942a.  Calculations were performed on the Uran cluster of
Ural branch RAS.

\setcounter{equation}{0}
\renewcommand\theequation{A\arabic{equation}}
\section*{Appendix. The transformation to even-odd states}

We first consider the case $t^{\alpha}_{2}=\eta t^{\alpha}_{1}, \alpha\in\{L,R\},0<\eta\le 1$, which includes up-down ($t^{L}_{1,2}=t^{R}_{1,2}$), left-right ($\eta=1, t^{L}_{1,2}\ne t^{R}_{1,2}$) asymmetry and perfect symmetry ($\eta=1, t^{L}_{1}=t^{R}_{1}$)  regimes of tunneling through the quantum dot system. In this case it is possible to define even-- ($d_{e,\sigma}$) and odd--parity ($d_{o,\sigma}$) orbitals:
\begin{equation}
\begin{pmatrix}
d_{e,\sigma}\\
d_{o,\sigma}\\
\end{pmatrix}=
\dfrac{1}{\sqrt{1+\eta^{2}}}
\begin{pmatrix}
1 & \eta\\
-\eta & 1\\
\end{pmatrix}
\begin{pmatrix}
d_{1,\sigma}\\
d_{2,\sigma}\\
\end{pmatrix}
\label{transf}
\end{equation}
in which only even--parity orbitals are directly connected to the leads and the coupling part of the Hamiltonian~(\ref{H_coupl}) takes the form:
\begin{equation}
\mathcal{H}_{\rm coupl}=-\sum_{\alpha=L,R}\sum_{\sigma}(t^{\alpha}_{e}c^{\dagger}_{\alpha,0,\sigma}d_{e,\sigma}+\text{H.c.}),
\end{equation}
where $t^{\alpha}_{e}=\sqrt{1+\eta^{2}}t^{\alpha}_{1}$. 

The dot part $\mathcal{H}_{dot}$ of the Hamiltonian~(\ref{H_dot}) in the even-odd basis can be represented as 
\begin{widetext}
\begin{eqnarray}
\mathcal{H}_{\rm dot}&=&\sum_{\sigma}\sum_{p\in\{e,o\}}\left(\epsilon_{\sigma}-\dfrac{U}{2}\right)n_{p,\sigma}-2J_{eo}\text{ }\vec{\mathbf{S}}_{e}\vec{\mathbf{S}}_{o}+\dfrac{U(1+\eta^{4})}{\left(1+\eta^{2}\right)^{2}}\left(n_{e,\uparrow}n_{e,\downarrow}+n_{o,\uparrow}n_{o,\downarrow}\right)+\dfrac{J_{eo}}{2}n_{e}n_{o} \notag\\  &+&J_{eo}\left(d^{\dagger}_{e,\uparrow}d_{o,\uparrow}d^{\dagger}_{e,\downarrow}d_{o,\downarrow}+\text{H.c.}\right)+ \dfrac{U\eta\left(1-\eta^{2}\right)}{\left(1+\eta^{2}\right)^{2}}\sum_{\sigma}\left(d^{\dagger}_{e,\sigma}d_{o,\sigma}+d^{\dagger}_{o,\sigma}d_{e,\sigma}\right)\left(n_{o,-\sigma}-n_{e,-\sigma}\right),\label{eq2}
\end{eqnarray}
\end{widetext}
where the particle number operators $n_{e/o,\sigma}$ and spin operators $\vec{\mathbf{S}}_{e/o}$ are defined as:
\begin{eqnarray}
n_{e/o}&=&\sum_{\sigma}n_{e/o,\sigma}=\sum_{\sigma}d^{\dagger}_{e/o,\sigma}d_{e/o,\sigma},\notag \\
\vec{\mathbf{S}}_{e/o}&=&\dfrac{1}{2}\sum_{\sigma,\sigma^{'}}d^{\dagger}_{e/o,\sigma}\vec{\bm{\sigma}}d_{e/o,\sigma^{'}}, 
\end{eqnarray}
here $\vec{\bm{\sigma}}$ are the Pauli matrices. Thus, the Hamiltonian~(\ref{H_dot}) can be mapped onto the two-orbital Hamiltonian, that includes the diagonal quadratic part (the first term), the standard Hund exchange inter--orbital interactions (the second term) with the exchange constant $J_{eo}={2U\eta^{2}}/{\left(1+\eta^{2}\right)^{2}}$, that has the maximum value $J^{max}_{eo}=U/2$ at the $\eta=1$; the density--density intra--orbital and inter--orbital interactions as well as pair hopping term (third to fifth term); and correlated hopping, which is generated due to asymmetry of the system (the last term) and absent in the symmetric case $\eta=1$.\par

After fRG approach is applied, due to the frequency independence of the vertices the initial quantum dot system can be viewed as the non-interacting one with the effective Hamiltonian
\begin{equation}
 \mathcal{H}_{\rm dot}^{\rm eff}= \sum_{j,\sigma}\epsilon_{j,\sigma}n_{j,\sigma}-\dfrac{1}{2}\sum_{j\neq j^{'},\sigma}\left(t^{\sigma}_{jj^{'}}d^{\dagger}_{j,\sigma}d_{j^{'},\sigma}+ \text{H.c.}\right),
\label{Heff}
\end{equation}
where $\epsilon_{j,\sigma}=\epsilon_{\sigma}+\Sigma^{\Lambda\rightarrow 0}_{jj,\sigma}$ are the renormalized energy levels of quantum dots (the term $U/2$ in Eq.~(\ref{H_dot}) is canceled by the contribution arising due to integration of the self-energy flow equation (see, Eq.~(\ref{Gamma})) from the scale $\Lambda=\infty$ to finite $\Lambda=\Lambda_{0}$) and $t^{\sigma}_{ij}=-\Sigma^{\Lambda\rightarrow 0}_{ij,\sigma}$ represents the renormalized inter-dot   hopping parameters.

Transformation of the Hamiltonian (\ref{Heff}) to the basis of the even- and odd-parity orbitals yields
\begin{equation}
 \mathcal{H}_{\rm dot}^{\rm eff}= \sum_{\sigma}\left[\left(\epsilon_{e,\sigma}n_{e,\sigma}+\epsilon_{o,\sigma}n_{o,\sigma}\right)-\left(t^{\sigma}_{eo}d^{\dagger}_{e,\sigma}d_{0,\sigma}+\text{H.c.}\right)\right]
\end{equation}
with the effective even and odd energy levels $\epsilon_{e/o,\sigma}$ and the effective hopping parameters $t^{\sigma}_{eo}$ are defined by the Eqs. (\ref{eeo}) and (\ref{Teo}) of the main text.  
The corresponding conductance for each spin projection can be represented in the 
form of 
Eq. (\ref{g_sigma}) of the main text.\par

It is important to note that from the explicit form of the fRG equations~(\ref{Gamma}), it follows that in the limit case of zero magnetic field $H\rightarrow 0$ the renormalized energy levels and hopping parameters satisfy the relations: $\epsilon_{e/o,\sigma}(V_{g})=-\epsilon_{e/o,-\sigma}(-V_{g})$ and $t_{eo}^{\sigma}(V_{g})=-t_{eo}^{-\sigma}(-V_{g})$, 
which allows us to write the total conductance of the system $g=g_{\uparrow}+g_{\downarrow}=g_{\sigma}(V_{g})+g_{\sigma}(-V_{g})$. Thus, the total conductance can be analyzed using the gate voltage dependence of $q_{\sigma}(\text{or }g_{\sigma})$ for only one spin projection.

\vspace{0.1cm}
In case of arbitrary asymmetry we can use the same transformation (\ref{transf}), however the coupling of the odd orbital to the leads does not vanish in general, and the corresponding part of the Hamiltonian takes the form
\begin{equation}
\mathcal{H}_{\rm coupl}=-\sum_{\alpha=L,R}\sum_{\sigma}(t^{\alpha}_{e}c^{\dagger}_{\alpha,0,\sigma}d_{e,\sigma}+t^{\alpha}_{o}c^{\dagger}_{\alpha,0,\sigma}d_{o,\sigma}+\text{H.c.})\notag
\end{equation}
with tunnel matrix elements $t^{\alpha}_{e}= at^{\alpha}_{1}+(1-a^{2})^{1/2}t^{\alpha}_{2}$ and $t^{\alpha}_{o}=a t^{\alpha}_{2}-(1-a^{2})^{1/2}t^{\alpha}_{1}$ where $a=(1+\eta^2)^{-1/2}$. The parameters $\eta$ and $a$ are determined in this case in the main text from the condition (\ref{cond1}).

\end{document}